\documentclass[12pt]{article}
\usepackage{amsmath,amssymb}
\usepackage{graphicx}
\newcommand{\eqdef}{\stackrel{\text{def}}{=}}
\newcommand{\n}{\nonumber\\}
\newcommand{\bm}{\boldsymbol}

\newcommand{\ait}{\text{II}}

\newcommand{\cF}{c_{\text{\tiny$\mathcal{F}$}}}
\newcommand{\ignore}[1]{}
\numberwithin{equation}{section}
\newcommand{\Romannumeral}[1]{\uppercase\expandafter{\romannumeral#1}}
\newcommand{\I}{\text{\Romannumeral{1}}}
\newcommand{\II}{\text{\Romannumeral{2}}}
\newcommand{\III}{\text{\Romannumeral{3}}}

\allowdisplaybreaks[4]

\begin{document}

\baselineskip=20pt

\newfont{\elevenmib}{cmmib10 scaled\magstep1}
\newcommand{\preprint}{
    \begin{flushleft}
     \elevenmib Yukawa\, Institute\, Kyoto\\
   \end{flushleft}\vspace{-1.3cm}
   \begin{flushright}\normalsize \sf
     DPSU-12-4\\
     YITP-12-96\\
   \end{flushright}}
\newcommand{\Title}[1]{{\baselineskip=26pt
   \begin{center} \Large \bf #1 \\ \ \\ \end{center}}}
\newcommand{\Author}{\begin{center}
   \large \bf Satoru Odake${}^a$ and Ryu Sasaki${}^b$ \end{center}}
\newcommand{\Address}{\begin{center}
     $^a$ Department of Physics, Shinshu University,\\
     Matsumoto 390-8621, Japan\\
     ${}^b$ Yukawa Institute for Theoretical Physics,\\
     Kyoto University, Kyoto 606-8502, Japan
   \end{center}}
\newcommand{\Accepted}[1]{\begin{center}
   {\large \sf #1}\\ \vspace{1mm}{\small \sf Accepted for Publication}
   \end{center}}

\preprint
\thispagestyle{empty}

\Title{Extensions of solvable potentials with finitely many discrete
eigenstates}

\Author

\Address
\vspace{1cm}

\begin{abstract}
We address the problem of rational extensions of six examples of
shape-invariant potentials having finitely many discrete eigenstates.
The {\em overshoot eigenfunctions\/} are proposed as candidates
unique in this group for the {\em virtual state wavefunctions\/}, which are an
essential ingredient for {\em multi-indexed\/} and {\em iso-spectral\/}
extensions of these potentials. They have exactly the same form as the
eigenfunctions but their degrees are much higher than $n_{\text{max}}$
so that their energies are lower than the groundstate energy.
\end{abstract}

\section{Introduction}
\label{intro}

Study of exactly solvable potentials in one dimensional quantum mechanics
is a very rapidly developing subject in theoretical/mathematical physics
in recent years.
The main focus is on solvable extensions (deformations) of known exactly
solvable potentials \cite{infhul, susyqm}, in particular, the radial
oscillator potential and the Darboux-P\"oschl-Teller potential.
These potentials have infinitely many discrete eigenstates and the
corresponding eigenfunctions consist of classical orthogonal polynomials,
the Laguerre and the Jacobi polynomials \cite{szego}. Their iso-spectral
extensions also produce complete sets of orthogonal polynomials,
generically known as multi-indexed Laguerre and Jacobi polynomials
\cite{os25, gomez3}--\cite{os27} and their various special cases;
the simplest single-indexed cases are called exceptional orthogonal
polynomials \cite{junkroy}--\cite{ramos2}.
The extensions (deformations) of these potentials are achieved by
Darboux-Crum transformations \cite{darb, crum} and its modification by
Krein-Adler \cite{adler}--\cite{gos} through the Wronskian formulas
\eqref{vDarb}--\eqref{vpot} in terms of {\em virtual state wavefunctions},
see the definition in section \ref{sec:vir}. In these cases, the virtual
state wavefunctions are obtained from the eigenfunctions by twisting
the parameters specified by the {\em discrete symmetries} of the
Hamiltonians (potentials), see (19)--(22) of \cite{os25}.
Thus these virtual state wavefunctions are also polynomial solutions.

In this paper we will formulate the general procedures of the extensions
(deformations) of solvable potentials with {\em finitely many discrete
eigenstates}, which are labeled by the degrees of the polynomial
eigenfunctions, $n=0,1,\ldots, n_{\text{max}}$.
We discuss six explicit examples; Morse potential (M) \S\ref{sec:M},
the soliton potential (s) \S\ref{sec:s}, Rosen-Morse potential (RM)
\S\ref{sec:RM}, the hyperbolic symmetric top \ait\ (hst) \S\ref{sec:hst},
Kepler problem in hyperbolic space (Eckart potential) (Kh) \S\ref{sec:Kh}
and the hyperbolic Darboux-P\"oschl-Teller potential (hDPT) \S\ref{sec:hDPT},
according to the naming of the review by  Infeld-Hull \cite{infhul}.
The essential point is again the identification of the virtual state
wavefunctions, since Darboux-Crum transformations apply equally well
in these cases.
The discrete symmetries of these potentials, except for (hst) and (hDPT),
produce only the {\em pseudo virtual state wavefunctions} and they provide
the known type of non iso-spectral extensions that are realised by the
deletion of eigenstates \`a la Krein-Adler \cite{adler} with shifted parameters
\cite{os29}--\cite{quesne6}.
This correspondence was proven in detail in \cite{os29} for the above
six potentials as well as other five well-known potentials having
infinitely many discrete eigenstates.

Good candidates for the {\em virtual states for those potentials having
only finitely many discrete eigenstates} are provided by the same
polynomial wavefunctions as the eigenfunctions with higher degrees than
the highest energy eigenfunction $n>n_{\text{max}}$, so that their
energies are lower than the ground state energy.
They are tentatively called {\em overshoot eigenfunctions}.
In her pioneering works \cite{quesne4, quesne5}, Quesne used overshoot
eigenfunctions for the ``first order SUSY'' (single-indexed) extensions
of Morse, Rosen-Morse and Eckart potentials.
This idea was further developed in \cite{grandati2}.
Once the virtual state wavefunctions are identified, the multi-indexed
extensions are trivially realised in the manner explained in \cite{os25}.
For some potentials {\em the overshoot eigenfunctions provide a new
type of pseudo virtual state wavefunctions}, which generate extensions
by adding one eigenstate below the groundstate.
These extensions are a new type and they cannot be realised by the deletion
of eigenstates \`a la Krein-Adler for generic parameters.
It can be easily seen that the energy of the added state does not
correspond to that of a {\em negative degree state} except for half
integer coupling constant cases.

Extended potentials in terms of virtual state wavefunctions do inherit
the shape-invariance \cite{genden} of the original potentials as in the
cases of the multi-indexed extensions \cite{os25}.
One new feature of the shape-invariance of the potentials with finitely
many discrete eigenstates (except for the hyperbolic Darboux-P\"oschl-Teller
potential) is that the degrees of the virtual state wavefunctions also
shift together with $n_{\text{max}}\to n_{\text{max}}-1$ as emphasised
in Quesne's papers \cite{quesne4,quesne5}.
Similar phenomena have been remarked in the extensions in terms of pseudo
virtual states \cite{os29}.

This paper is organised as follows.
The concepts of the virtual and pseudo virtual state wavefunctions are
explained briefly in section \ref{sec:vir} and the Darboux-Crum
transformation formulas in terms of multiple (pseudo) virtual state
wavefunctions are recapitulated. The idea of the overshoot eigenfunctions is
introduced in section \ref{sec:over} together with the illustration of
three types of the energy curves of the systems having finitely many
discrete eigenstates.
Section \ref{sec:Ex} is the main part of the paper. Various examples of
the virtual state wavefunctions and the new type of pseudo virtual state wave
functions obtained from the overshoot eigenfunctions are discussed in
some detail for the six potentials having finitely many discrete
eigenstates.
The final section is for a summary and comments.

\section{Virtual and Pseudo Virtual States}
\label{sec:vir}

Here we recapitulate the concepts of virtual state and pseudo virtual
state wavefunctions. For more details on the extensions of solvable
potentials  through Darboux-Crum and Krein-Adler transformations in
terms of virtual state and pseudo virtual state wavefunctions,
we refer to the recent publication \cite{os29}.

We consider quantum mechanical systems defined in an interval $x_1<x<x_2$,
which have a finite number of discrete eigenstates with a vanishing
groundstate energy,
\begin{align}
  &\mathcal{H}\phi_n(x)=\mathcal{E}_n\phi_n(x)
  \ \ (n=0,1,\ldots,n_{\text{max}}),\quad
  0=\mathcal{E}_0<\mathcal{E}_1<\cdots<\mathcal{E}_{n_{\text{max}}},
  \label{sheq}\\
  &(\phi_m,\phi_n)\eqdef\int_{x_1}^{x_2}\!dx\,\phi_m(x)\phi_n(x)
  =h_n\delta_{m\,n}\quad(h_n>0).
  \label{inpro}
\end{align}
The main part of $\phi_n$ is a certain polynomial.
Here are the conditions for the {\em virtual state wavefunctions}
\footnote{
They are different from the `virtual energy levels' in quantum scattering
theory \cite{schiff}.}
$\tilde{\phi}(x)$, $\mathcal{H}\tilde{\phi}(x)
=\tilde{\mathcal{E}}\tilde{\phi}(x)$:
\begin{enumerate}
\item No zeros in $x_1<x<x_2$, {\em i.e.} $\tilde{\phi}(x)>0$ or
$\tilde{\phi}(x)<0$ in $x_1<x<x_2$.
\item Negative energy, $\tilde{\mathcal{E}}<0$.
\item $\tilde{\phi}$ is also a polynomial type solution,
like the original eigenfunctions, see \eqref{gra},\eqref{grb}.
\item Square non-integrability, $(\tilde{\phi},\tilde{\phi})=\infty$.
\item Reciprocal square non-integrability,
$(\tilde{\phi}^{-1},\tilde{\phi}^{-1})=\infty$.
\end{enumerate}
Of course these conditions are not totally independent.
The negative energy condition is necessary for the positivity of the
norm as seen from the norm formula \eqref{Mnorm2}.
When the first condition is dropped and the reciprocal is required
to be square integrable at both boundaries, $(x_1,x_1+\epsilon)$,
$(x_2-\epsilon,x_2)$, $\epsilon>0$, see \eqref{type3},
such seed functions are called {\em pseudo virtual state wavefunctions}.
When the system is extended in terms of a pseudo virtual state wavefunction
$\tilde{\phi}(x)$, the new Hamiltonian has an extra {\em eigenstate}
$\tilde{\phi}^{-1}(x)$ with the eigenvalue $\tilde{\mathcal{E}}$,
{\em if the new potential is non-singular}.
The extra state is below the original groundstate and the extension is
no longer iso-spectral.
This is a well known result of the Darboux transformation \cite{darb},
\cite{sukumar}.
Since the (pseudo) virtual state wavefunctions $\tilde{\phi}(x)$ are finite
in $x_1<x<x_2$, the non-square integrability can only be caused by the
boundaries. Thus the virtual state wavefunctions belong to either of the
following type $\I$ and $\II$ and the pseudo virtual state wavefunctions
belong to type $\III$ :
\begin{alignat}{3}
  \text{Type $\I$}:&&
  &\int_{x_1}^{x_1+\epsilon}\!\!\!dx\,\tilde{\phi}(x)^2<\infty,\quad
  &&\int_{x_2-\epsilon}^{x_2}\!\!\!dx\,\tilde{\phi}(x)^2=\infty,\n
  &\ \ \text{\&}\ &&\int_{x_1}^{x_1+\epsilon}\!\!\!dx\,
  \tilde{\phi}(x)^{-2}=\infty,\quad
  &&\int_{x_2-\epsilon}^{x_2}\!\!\!dx\,\tilde{\phi}(x)^{-2}<\infty,
  \label{type1}\\
  \text{Type $\II$}:&&
  &\int_{x_1}^{x_1+\epsilon}\!\!\!dx\,\tilde{\phi}(x)^2=\infty,\quad
  &&\int_{x_2-\epsilon}^{x_2}\!\!\!dx\,\tilde{\phi}(x)^2<\infty,\n
  &\ \ \text{\&}\ &&\int_{x_1}^{x_1+\epsilon}\!\!\!dx\,
  \tilde{\phi}(x)^{-2}<\infty,\quad
  &&\int_{x_2-\epsilon}^{x_2}\!\!\!dx\,\tilde{\phi}(x)^{-2}=\infty,
  \label{type2}\\
  \text{Type $\III$}:&&
  &\int_{x_1}^{x_1+\epsilon}\!\!\!dx\,\tilde{\phi}(x)^2=\infty\ \ \text{or}
  &&\int_{x_2-\epsilon}^{x_2}\!\!\!dx\,\tilde{\phi}(x)^2=\infty,\n
  &\ \ \text{\&}\ &&\int_{x_1}^{x_1+\epsilon}\!\!\!dx\,
  \tilde{\phi}(x)^{-2}<\infty,\quad
  &&\int_{x_2-\epsilon}^{x_2}\!\!\!dx\,\tilde{\phi}(x)^{-2}<\infty.
  \label{type3}
\end{alignat}
An appropriate modification is needed when $x_2=+\infty$ and/or $x_1=-\infty$.

The Darboux-Crum transformations in terms of
$\mathcal{D}\eqdef\{d_1,d_2,\ldots,d_M\}$ (pseudo) virtual state
wavefunctions ($\mathcal{H}\tilde{\phi}_{\text{v}}(x)
=\tilde{\mathcal{E}}_{\text{v}}\tilde{\phi}_{\text{v}}(x)$) read:
\begin{align}
  &\mathcal{H}^{[M]}\phi_n^{[M]}(x)=\mathcal{E}_n\phi_n^{[M]}(x),
  \label{vDarb}\\
  &\phi_n^{[M]}(x)\eqdef
  \frac{\text{W}[\tilde{\phi}_{d_1},\tilde{\phi}_{d_2},\ldots,
  \tilde{\phi}_{d_M},\phi_n](x)}
  {\text{W}[\tilde{\phi}_{d_1},\tilde{\phi}_{d_2},\ldots,
  \tilde{\phi}_{d_M}](x)},\quad
  ({\phi}_m^{[M]},\phi_n^{[M]})
  =\prod_{j=1}^M(\mathcal{E}_n-\tilde{\mathcal E}_{d_j})
  \cdot h_n\delta_{m\,n},
  \label{Mnorm2}\\
  &U^{[M]}(x)\eqdef U(x)-2\partial_x^2\log
  \bigl|\text{W}[\tilde{\phi}_{d_1},\tilde{\phi}_{d_2},\ldots,
  \tilde{\phi}_{d_M}](x)\bigr|,
  \label{vpot}\\
  &\mathcal{H}^{[M]}\tilde{\phi}_{\text{v}}^{[M]}(x)
  =\tilde{\mathcal{E}}_{\text{v}}\tilde{\phi}_{\text{v}}^{[M]}(x),\quad
  \tilde{\phi}^{[M]}_{\text{v}}(x)
  =\frac{\text{W}[\tilde{\phi}_{d_1},\tilde{\phi}_{d_2},\ldots,
  \tilde{\phi}_{d_M},\tilde{\phi}_{\text{v}}](x)}
  {\text{W}[\tilde{\phi}_{d_1},\tilde{\phi}_{d_2},\ldots,
  \tilde{\phi}_{d_M}](x)}.
  \label{tphiMv}
\end{align}%
In \eqref{Mnorm2} $h_n$ is the normalisation constant of the original system
defined in \eqref{inpro}.
Note that the differential equations \eqref{vDarb} and \eqref{tphiMv}
hold irrespective of the presence of the nodes of the Wronskian
$\text{W}[\tilde{\phi}_{d_1},\tilde{\phi}_{d_2},\ldots,\tilde{\phi}_{d_M}](x)$.
By using the Wronskian identity
\begin{align*}
  &\text{W}\bigl[\text{W}[f_1,f_2,\ldots,f_n,g],
  \text{W}[f_1,f_2,\ldots,f_n,h]\,\bigr](x)\n
  &=\text{W}[f_1,f_2,\ldots,f_n](x)\cdot
  \text{W}[f_1,f_2,\ldots,f_n,g,h](x)
  \quad(n\geq 0),
\end{align*}
and \eqref{tphiMv}, we can show the following ($s\geq 1$):
\begin{align}
  &\quad\frac{d}{dx}
  \frac{\text{W}[\tilde{\phi}_{d_1},\tilde{\phi}_{d_2},\ldots,
  \tilde{\phi}_{d_{s-1}},\tilde{\phi}_{d_s},\tilde{\phi}_{d_{s+1}}](x)}
  {\text{W}[\tilde{\phi}_{d_1},\tilde{\phi}_{d_2},\ldots,
  \tilde{\phi}_{d_{s-1}}](x)}\n
  &=(\tilde{\mathcal{E}}_{d_s}-\tilde{\mathcal{E}}_{d_{s+1}})
  \frac{\text{W}[\tilde{\phi}_{d_1},\tilde{\phi}_{d_2},\ldots,
  \tilde{\phi}_{d_{s-1}},\tilde{\phi}_{d_s}](x)}
  {\text{W}[\tilde{\phi}_{d_1},\tilde{\phi}_{d_2},\ldots,
  \tilde{\phi}_{d_{s-1}}](x)}
  \frac{\text{W}[\tilde{\phi}_{d_1},\tilde{\phi}_{d_2},\ldots,
  \tilde{\phi}_{d_{s-1}},\tilde{\phi}_{d_{s+1}}](x)}
  {\text{W}[\tilde{\phi}_{d_1},\tilde{\phi}_{d_2},\ldots,
  \tilde{\phi}_{d_{s-1}}](x)}.
  \label{ddxW}
\end{align}
For deletion of type $\I$ virtual sate wavefunctions only (or type $\II$ only),
equation \eqref{ddxW} implies inductively that
$\text{W}[\tilde{\phi}_{d_1},\tilde{\phi}_{d_2},\ldots,\tilde{\phi}_{d_M}](x)$
has no node in $x_1<x<x_2$ under the conditions
\begin{equation}
 \partial_x^s\tilde{\phi}_{d_j}(x)\bigm|_{x=\text{$x_1$ (or $x_2$)}}=0
 \ \ (s=0,1,\ldots,M-1).
\end{equation}
For deletion of both type $\I$ and type $\II$ virtual state wavefunctions
such as in \cite{os25} and \S\,\ref{sec:hDPT},
we have to check that 
$\text{W}[\tilde{\phi}_{d_1},\tilde{\phi}_{d_2},\ldots,
\tilde{\phi}_{d_{s-1}},\tilde{\phi}_{d_s},\tilde{\phi}_{d_{s+1}}](x)/
\text{W}[\tilde{\phi}_{d_1},\tilde{\phi}_{d_2},\ldots,
\tilde{\phi}_{d_{s-1}}](x)$ has the same sign at $x=x_1$ and $x=x_2$,
for no-nodeness of 
$\text{W}[\tilde{\phi}_{d_1},\tilde{\phi}_{d_2},\ldots,\tilde{\phi}_{d_M}](x)$.

\section{Overshoot Eigenfunctions}
\label{sec:over}

The simplest way to construct the virtual and pseudo virtual wavefunctions
is to twist the parameters of the eigenfunctions based on the discrete
symmetries of the original shape-invariant Hamiltonians.
This automatically guarantee that the obtained (pseudo) virtual
wavefunctions are of the same polynomial type as the original eigenfunctions.
Among the six examples listed in \S\,\ref{sec:Ex}, the discrete symmetries
produce only pseudo virtual state wavefunctions except for the well-known cases
of the hyperbolic DPT \cite{os18,hos,os25,os29} and the Eckart potential
(Kh) \cite{quesne5,os29}. 

Here we propose {\em overshoot eigenfunctions\/}
$\tilde{\phi}^{\text{os}}_{\text{v}}(x)$
as the candidates for new types of virtual state and pseudo virtual
eigenfunctions
for the six examples of shape-invariant potentials listed in \S\,\ref{sec:Ex}.
The overshoot eigenfunctions have exactly the same forms as the eigenfunctions
$\tilde{\phi}^{\text{os}}_{\text{v}}(x)\eqdef\phi_{\text{v}}(x)$
($\tilde{\mathcal{E}}^{\text{os}}_{\text{v}}=\mathcal{E}_{\text{v}}$).
But their degrees are much higher than the highest discrete energy level
$n_{\text{max}}$ so that their energies are lower than the groundstate energy,
which are chosen zero throughout this paper.
The following illustrations of the three types of energy curves
$\mathcal{E}_n$ with respect to the degree $n$ would be helpful for
understanding (Figure 1).
In all the three cases the region (a) corresponds to the finitely many discrete
eigenstates. The overshoot eigenfunctions correspond to the region (b),
which provides the candidates of virtual state wavefunctions and a new
type of pseudo virtual wavefunctions. The region (c) in the first two
illustrations and the region (c$_1$) in the third corresponds to the pseudo
virtual state wavefunctions.

\begin{figure}[htbp]
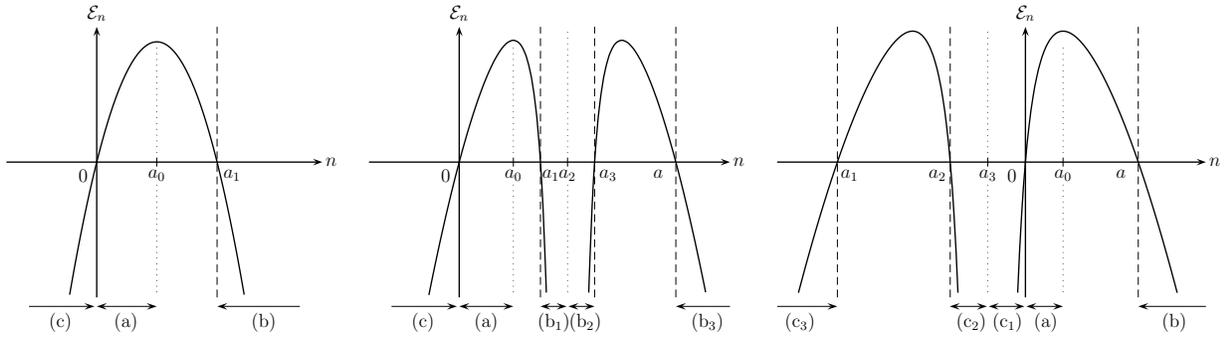

\begin{center}
  \scalebox{0.6}{\includegraphics{en1.epsi}}\quad
  \scalebox{0.6}{\includegraphics{en2.epsi}}\quad
  \scalebox{0.6}{\includegraphics{en3.epsi}}
\caption{The left represents the energy levels of (M), (s), (hst) and (hDPT).
The center corresponds to (RM) and the right to (Kh).}
\end{center}
\end{figure}

\section{Examples of (Pseudo) Virtual State Wavefunctions}
\label{sec:Ex}

Here we provide the explicit forms of the virtual state wavefunctions
and a new type of pseudo virtual wavefunctions obtained from the
{\em overshoot eigenfunctions} for six shape-invariant systems
\S\ref{sec:M}--\S\ref{sec:hDPT}, which have finitely many discrete eigenstates. 
The original systems to be extended usually contain some parameter(s),
$\bm{\lambda}=(\lambda_1,\lambda_2,\ldots)$ and the parameter dependence
is denoted by $\mathcal{H}(\bm{\lambda})$, $\mathcal{E}_n(\bm{\lambda})$,
$\phi_n(x;\bm{\lambda})$,  $\tilde{\phi}_n(x;\bm{\lambda})$,
$P_n(\eta(x);\bm{\lambda})$ etc.
As discussed in \cite{os29} they are divided into two groups of
{\em eigenfunction patterns\/}:
\begin{alignat}{2}
  &\text{Group A: (M),\,(s),\,(hst),\,(hDPT)},
  &\ \ &\phi_n(x;\bm{\lambda})
  =\phi_0(x;\bm{\lambda})P_n\bigl(\eta(x);\bm{\lambda}\bigr),
  \label{gra}\\
  &\text{Group B: (RM),\,(Kh)},
  &\ \ &\phi_n(x;\bm{\lambda})
  =\phi_0(x;\bm{\lambda}+n\bm{\delta})P_n\bigl(\eta(x);\bm{\lambda}\bigr).
  \label{grb}
\end{alignat}
Here $P_n\bigl(\eta(x);\bm{\lambda}\bigr)$ is a polynomial of degree $n$
in a certain function $\eta(x)$, which is called the sinusoidal coordinate
(for Group A) \cite{os7}.
Except for (Kh) and (hDPT), the twisting of
the parameter(s) of the eigenfunctions based on the discrete symmetries
of the Hamiltonian generate pseudo virtual states, rather than the
virtual states.
We will not discuss the discrete symmetries except for (Kh) and (hDPT),
since the pseudo virtual states and the rational extensions of these six
potentials in terms of them have been explored in detail in \cite{os29}.
For the parameters of shape-invariant transformation
$\bm{\lambda}\to\bm{\lambda}+\bm{\delta}$, we use the symbol $g$ to
denote an  increasing ($\bm{\delta}=1$) parameter and the symbol $h$
for a decreasing ($\bm{\delta}=-1$) parameter and $\mu$ for an unchanging
($\bm{\delta}=0$) parameter.
Except for \S\,\ref{sec:halfint}, we assume that the parameter $g$ and/or
$h$ take {\em generic values\/}, that is not integers or half odd integers.
We refer to a recent paper \cite{os29} for various properties of these
six potentials.
No-nodeness of the virtual state wavefunctions can be verified by using
the properties of zeros of Laguerre and Jacobi polynomials
\cite{szego,grandati,ho1,quesne5}.

The energy of a {\em new type of pseudo virtual state\/}
$\tilde{\phi}^{\text{os}}_{\text v}(x)$ obtained from the overshoot
eigenfunctions does not coincide with the negative energy levels of the
original theory except for half integer coupling constant cases.
In other words, the previous formula
$\tilde{\mathcal{E}}_{\text{v}}(\bm{\lambda})
=\mathcal{E}_{-\text{v}-1}(\bm{\lambda})$ in \cite{os29} is no longer valid
for generic coupling constants.
This means that the extensions in terms of the new type of pseudo virtual
states cannot be reproduced in terms of the Krein-Adler transformations
deleting the (multiple) eigenstates except for half integer coupling
constant cases.
In other words, the extensions in terms of one pseudo virtual wavefunction
of the new type can be made non-singular for some choices of the degrees.
But the extensions by using two or more such pseudo virtual wavefunctions
would produce singular potentials, since the nodeless condition of the seed
functions is no longer guaranteed.

The Wronskian of the eigenfunctions $\phi_n(x;\bm{\lambda})$ is expressed
as \cite{os29}
\begin{align}
  &\text{W}[\phi_{d_1},\phi_{d_2},\ldots,\phi_{d_M}](x;\bm{\lambda})
  =\bar{A}_{\mathcal{D}}(x;\bm{\lambda})
  \bar{\Xi}_{\mathcal{D}}\bigl(\eta(x);\bm{\lambda}\bigr),
  \label{Wphi=AXi}\\
  &\bar{A}_{\mathcal{D}}(x;\bm{\lambda})\eqdef\left\{
  \begin{array}{ll}
  \phi_0(x;\bm{\lambda})^M
  \bigl(\cF^{-1}\frac{d\eta(x)}{dx}\bigr)^{\frac12M(M-1)}
  &:\text{Group A}\\[4pt]
  \prod_{k=1}^M\phi_0(x;\bm{\lambda}+d_k\bm{\delta})
  &:\text{Group B}
  \end{array}\right.,\\
  &\bar{\Xi}_{\mathcal{D}}(\eta;\bm{\lambda})\eqdef\left\{
  \begin{array}{ll}
  \cF^{\,\frac12M(M-1)}
  \text{W}[P_{d_1},P_{d_2},\ldots,P_{d_M}](\eta;\bm{\lambda})
  &:\text{Group A}\\[4pt]
  \det\bigl(\bar{X}^{\mathcal{D}}_{j,k}(\eta;\bm{\lambda})
  \bigr)_{1\leq j,k\leq M}
  &:\text{Group B}
  \end{array}\right.,\\
  &\bar{X}^{\mathcal{D}}_{j,k}(\eta;\bm{\lambda})\eqdef
  \prod_{i=0}^{j-2}f_{d_k-i}(\bm{\lambda}+i\bm{\delta})\cdot
  P_{d_k-j+1}\bigl(\eta;\bm{\lambda}+(j-1)\bm{\delta}\bigr),
\end{align}
where $\cF$ is $-1$ (M), $1$ (s), $1$ (hst) and $4$ (hDPT)
(See \underline{Remark} below (3.27) in \cite{os29}).
The denominator polynomial $\bar{\Xi}_{\mathcal{D}}(\eta;\bm{\lambda})$ is
a polynomial in $\eta$ and its degree is generically $\ell_{\mathcal{D}}$:
\begin{equation}
  \ell_{\mathcal{D}}\eqdef\sum_{j=1}^Md_j-\frac12M(M-1).
  \label{ellD}
\end{equation}
The ratio of the Wronskians becomes
\begin{align}
  &\frac{\text{W}[\phi_{d_1},\ldots,\phi_{d_M},\phi_n](x;\bm{\lambda})}
  {\text{W}[\phi_{d_1},\ldots,\phi_{d_M}](x;\bm{\lambda})}
  =\frac{\bar{P}_{\mathcal{D},n}\bigl(\eta(x);\bm{\lambda}\bigr)}
  {\bar{\Xi}_{\mathcal{D}}\bigl(\eta(x);\bm{\lambda}\bigr)}
  \times\left\{
  \begin{array}{ll}
  \phi_0(x;\bm{\lambda}+M\bm{\delta})
  &:\text{Group A}\\[2pt]
  \phi_0(x;\bm{\lambda}+n\bm{\delta})
  &:\text{Group B}
  \end{array}\right.,\n
  &\bar{P}_{\mathcal{D},n}(\eta;\bm{\lambda})\eqdef
  \bar{\Xi}_{\{d_1,\ldots,d_M,n\}}(\eta;\bm{\lambda})
  \ \ :\ \text{degree $=\ell_{\{d_1,\ldots,d_M,n\}}=~\ell_{\mathcal{D}}-M+n$}.
\end{align}
Since these expressions of the Wronskians are algebraic, they are applicable
to the overshoot eigenfunctions
$\tilde{\phi}^{\text{os}}_{\text{v}}(x;\bm{\lambda})
=\phi_{\text{v}}(x;\bm{\lambda})$
($\tilde{\mathcal{E}}^{\text{os}}_{\text{v}}(\bm{\lambda})
=\mathcal{E}_{\text{v}}(\bm{\lambda})$).

\subsection{Morse potential (M)}
\label{sec:M}

The system has finitely many discrete eigenstates
$0\le n\le n_\text{max}(\bm{\lambda})=[h]'$ in the specified parameter range
($[a]'$ denotes the greatest integer not exceeding and not equal to $a$):
\begin{align*}
  &\bm{\lambda}=(h,\mu),\quad\bm{\delta}=(-1,0),\quad
  -\infty<x<\infty,\quad h,\mu>0,\\
  &w(x;\bm{\lambda})=hx-\mu e^x,\quad
  U(x;\bm{\lambda})=\mu^2e^{2x}-\mu(2h+1)e^x+h^2,\\
  &\mathcal{E}_n(\bm{\lambda})=h^2-(h-n)^2,\quad\eta(x)=e^{-x},\quad
  f_n(\bm{\lambda})=\frac{n-2h}{2\mu},\quad
  b_{n-1}(\bm{\lambda})=-2n\mu,\\
  &\phi_n(x;\bm{\lambda})
  =\phi_0(x;\bm{\lambda})P_n\bigl(\eta(x);\bm{\lambda}\bigr),\quad
  \phi_0(x;\bm{\lambda})=e^{hx-\mu e^x},\\
  &P_n(\eta;\bm{\lambda})=(2\mu\eta^{-1})^{-n}L_n^{(2h-2n)}(2\mu\eta^{-1}),
  \quad h_n(\bm{\lambda})=\frac{\Gamma(2h-n+1)}{(2\mu)^{2h}n!\,2(h-n)}.
\end{align*}

\subsubsection{virtual states}

The energy spectrum is depicted in the first graphic
($\mathcal{E}_n(\bm{\lambda})<0\ (n\geq 0)\Leftrightarrow n>2h$) of Fig. 1.
The overshoot eigenfunctions provide type $\II$ virtual state wavefunctions
for $\text{v}>2h$ \cite{quesne5}:
\begin{equation}
  \tilde{\phi}^{\text{os}}_{\text{v}}(x;\bm{\lambda})
  =\phi_{\text{v}}(x;\bm{\lambda}),\quad
  \tilde{\mathcal{E}}^{\text{os}}_{\text{v}}(\bm{\lambda})
  =\text{v}(2h-\text{v})\quad(\text{v}>2h).
\end{equation}
Since this type $\II$ virtual state wavefunction satisfies the boundary
conditions
\begin{equation}
  \partial_x^s\tilde{\phi}^{\text{os}}_{\text{v}}(x;\bm{\lambda})
  \bigm|_{x=\infty}=0\quad(s=0,1,\ldots),
\end{equation}
a multiple virtual state $\tilde{\phi}^{\text{os}}_{\text{v}}$ deletion gives
a non-singular Hamiltonian $\mathcal{H}^{[M]}$.
Their `shape-invariance' relation is
\begin{align}
  &w_{\mathcal{D}}(x;\bm{\lambda})
  \eqdef\log\biggl|
  \frac{\text{W}[\tilde{\phi}^{\text{os}}_{d_1},\ldots,
  \tilde{\phi}^{\text{os}}_{d_M},\phi_0](x;\bm{\lambda})}
  {\text{W}[\tilde{\phi}^{\text{os}}_{d_1},\ldots,
  \tilde{\phi}^{\text{os}}_{d_M}](x;\bm{\lambda})}\biggr|
  \quad(\min_j{d_j}\geq 2),\n
  &\bigl(\partial_xw_{\mathcal{D}}(x;\bm{\lambda})\bigr)^2
  -\partial_x^2w_{\mathcal{D}}(x;\bm{\lambda})
  =\bigl(\partial_xw_{\mathcal{D}_-}
  (x;\bm{\lambda}+\bm{\delta})\bigr)^2
  +\partial_x^2w_{\mathcal{D}_-}(x;\bm{\lambda}+\bm{\delta})
  +\mathcal{E}_1(\bm{\lambda}),
  \label{wDsi-}
\end{align}
where $\mathcal{D}_-\eqdef\{d_1-1,d_2-1,\ldots,d_M-1\}$ \cite{os29}.

\subsection{Soliton potential (s)}
\label{sec:s}

The system has finitely many discrete eigenstates
$0\le n\le n_\text{max}(\bm{\lambda})=[h]'$ in the specified parameter range:
\begin{align*}
  &\bm{\lambda}=h,\quad\bm{\delta}=-1,\quad
  -\infty<x<\infty,\quad h>0,\\
  &w(x;\bm{\lambda})=-h\log\cosh x,\quad
  U(x;\bm{\lambda})=-\frac{h(h+1)}{\cosh^2x}+h^2,\\
  &\mathcal{E}_n(\bm{\lambda})=h^2-(h-n)^2,\quad\eta(x)=\sinh x,\quad
  f_n(\bm{\lambda})=h,\quad b_{n-1}(\bm{\lambda})=\frac{n(2h-n)}{h},\\
  &\phi_n(x;\bm{\lambda})
  =\phi_0(x;\bm{\lambda})P_n\bigl(\eta(x);\bm{\lambda}\bigr),\quad
  \phi_0(x;\bm{\lambda})=(\cosh x)^{-h},\\
  &P_n\bigl(\eta(x);\bm{\lambda}\bigr)=(\cosh x)^nP_n^{(h-n,h-n)}(\tanh x),
  \quad
  h_n(\bm{\lambda})=\frac{2^{2h-2n}\Gamma(h+1)^2}{n!\,(h-n)\Gamma(2h-n+1)}.
\end{align*}
One can rewrite $P_n(\eta;\bm{\lambda})$ as
\begin{equation*}
  P_n(\eta;\bm{\lambda})=
  \frac{(h-[\frac{n-1}{2}])_{[\frac{n+1}{2}]}}{(h-n+\frac12)_{[\frac{n+1}{2}]}}
  i^nP_n^{(-h-\frac12,-h-\frac12)}(i\eta),
\end{equation*}
where $[a]$ denotes the greatest integer not exceeding $a$.
Note that if we take $\eta(x)=\tanh x$, the eigenfunction has the form
of Group B,
$\phi_n(x;\bm{\lambda})=(\cosh x)^{-h+n}\times P^{(h-n,h-n)}_n(\tanh x)$.

\subsubsection{pseudo virtual states}

The energy spectrum is depicted in the first graphic
($\mathcal{E}_n(\bm{\lambda})<0\ (n\geq 0)\Leftrightarrow n>2h$) of Fig. 1.
The overshoot eigenfunctions provide a new type of pseudo virtual state
wavefunctions for $\text{v}>2h$:
\begin{equation}
  \tilde{\phi}^{\text{os}}_{\text{v}}(x;\bm{\lambda})
  =\phi_{\text{v}}(x;\bm{\lambda}),\quad
  \tilde{\mathcal{E}}^{\text{os}}_{\text{v}}(\bm{\lambda})
  =\text{v}(2h-\text{v})\quad(\text{v}>2h).
\end{equation}
In the absence of virtual state wavefunctions, shape-invariant rational
extensions of the soliton potential are not possible.

\subsection{Rosen-Morse potential (RM)}
\label{sec:RM}

This potential is also called Rosen-Morse $\II$ potential.
The system has finitely many discrete eigenstates
$0\le n\le n_\text{max}(\bm{\lambda})=[h-\sqrt{\mu}\,]'$ in the specified
parameter range:
\begin{align*}
  &\bm{\lambda}=(h,\mu),\quad\bm{\delta}=(-1,0),\quad
  -\infty<x<\infty,\quad h>\sqrt{\mu}>0,\\
  &w(x;\bm{\lambda})=-h\log\cosh x-\frac{\mu}{h}x,\quad
  U(x;\bm{\lambda})=-\frac{h(h+1)}{\cosh^2x}+2\mu\tanh x
  +h^2+\frac{\mu^2}{h^2},\\
  &\mathcal{E}_n(\bm{\lambda})=h^2-(h-n)^2
  +\frac{\mu^2}{h^2}-\frac{\mu^2}{(h-n)^2},\quad\eta(x)=\tanh x,\n
  &f_n(\bm{\lambda})=\frac{h^2(h-n)^2-\mu^2}{h(h-n)^2},\quad
  b_{n-1}(\bm{\lambda})=\frac{n(2h-n)}{h},\\
  &\phi_n(x;\bm{\lambda})=e^{-\frac{\mu}{h-n}x}(\cosh x)^{-h+n}
  P_n\bigl(\eta(x);\bm{\lambda}\bigr),\quad 
  \phi_0(x;\bm{\lambda})=e^{-\frac{\mu}{h}x}(\cosh x)^{-h},\n
  &P_n(\eta;\bm{\lambda})=P_n^{(\alpha_n,\beta_n)}(\eta),\quad
  \alpha_n=h-n+\frac{\mu}{h-n},\ \ \beta_n=h-n-\frac{\mu}{h-n},\\
  &h_n(\bm{\lambda})=\frac{2^{2h-2n}(h-n)
  \Gamma(h+\frac{\mu}{h-n}+1)\Gamma(h-\frac{\mu}{h-n}+1)}
  {n!\,\bigl((h-n)^2-\frac{\mu^2}{(h-n)^2}\bigr)\Gamma(2h-n+1)}.
\end{align*}

\subsubsection{virtual states and pseudo virtual states}

The energy spectrum is depicted in the second graphic
($\mathcal{E}_n(\bm{\lambda})<0\ (n\geq 0)\Leftrightarrow
h-\frac{\mu}{h}<n<h$ or $h<n<h+\frac{\mu}{h}$ or $n>2h$) of Fig. 1.
The overshoot eigenfunctions provide two types of virtual state wavefunctions
\cite{quesne5} and a new type of pseudo virtual wavefunctions:
\begin{align}
  &\tilde{\phi}^{\text{os}}_{\text{v}}(x;\bm{\lambda})
  =\phi_{\text{v}}(x;\bm{\lambda}),\quad
  \tilde{\mathcal{E}}^{\text{os}}_{\text{v}}(\bm{\lambda})
  =\text{v}(2h-\text{v})\frac{(h-\text{v}+\frac{\mu}{h})
  (h-\text{v}-\frac{\mu}{h})}{(h-\text{v})^2},\n
  &\qquad
  \left\{\begin{array}{lll}
  \text{type $\II$ virtual states}&:&
  {\displaystyle h-\frac{\mu}{h}<\text{v}<h}\\
  \text{type $\I$ \ virtual states}&:&
  {\displaystyle h<\text{v}<h+\frac{\mu}{h}}\\
  \text{pseudo virtual states}&:&
  {\displaystyle \text{v}>2h}
  \end{array}\right..
\end{align}
The type II virtual states correspond to the region (b$_1$), whereas the
overshoot eigenfunctions in region (b$_2$) provide type I virtual states.
The negative energy condition $\tilde{\mathcal{E}}_{\text{v}}<0$ is
also satisfied in the region (b$_3$), $\text{v}>2h$, and in this range
the overshoot eigenfunctions provide the new type of pseudo virtual state
wavefunctions.
Since these virtual state wavefunctions satisfy the boundary conditions
\begin{align}
  \text{type $\II$ virtual state}:\quad
  &\partial_x^s\tilde{\phi}^{\text{os}}_{\text{v}}(x;\bm{\lambda})
  \bigm|_{x=\infty}=0\quad(s=0,1,\ldots),\\
  \text{type $\I$ virtual state}:\quad
  &\partial_x^s\tilde{\phi}^{\text{os}}_{\text{v}}(x;\bm{\lambda})
  \bigm|_{x=-\infty}=0\quad(s=0,1,\ldots),
\end{align}
the multiple virtual state $\tilde{\phi}^{\text{os}}_{\text{v}}$ deletion
(type $\II$ only or type $\I$ only)
gives a non-singular Hamiltonian $\mathcal{H}^{[M]}$.
They satisfy the `shape-invariance' relation \eqref{wDsi-}.

\subsection{Hyperbolic symmetric top $\II$ (hst)}
\label{sec:hst}

The system has finitely many discrete eigenstates
$0\le n\le n_\text{max}(\bm{\lambda})=[h]'$ in the specified parameter range:
\begin{align*}
  &\bm{\lambda}=(h,\mu),\quad\bm{\delta}=(-1,0),\quad
  -\infty<x<\infty,\quad h,\mu>0,\\
  &w(x;\bm{\lambda})=-h\log\cosh x-\mu\tan^{-1}\sinh x,\n
  &U(x;\bm{\lambda})=\frac{-h(h+1)+\mu^2+\mu(2h+1)\sinh x}{\cosh^2x}+h^2,\\
  &\mathcal{E}_n(\bm{\lambda})=h^2-(h-n)^2,\quad\eta(x)=\sinh x,\quad
  f_n(\bm{\lambda})=\frac{n-2h}{2},\quad b_{n-1}(\bm{\lambda})=-2n,\\
  &\phi_n(x;\bm{\lambda})
  =\phi_0(x;\bm{\lambda})
  P_n\bigl(\eta(x);\bm{\lambda}\bigr),\quad
  \phi_0(x;\bm{\lambda})=e^{-\mu\tan^{-1}\sinh x}(\cosh x)^{-h},\n
  &P_n(\eta;\bm{\lambda})=i^{-n}P_n^{(\alpha,\beta)}(i\eta),\quad
  \alpha=-h-\tfrac12-i\mu,\ \ \beta=-h-\tfrac12+i\mu,\\
  &h_n(\bm{\lambda})=\frac{\pi\Gamma(2h-n+1)}
  {2^{2h}n!\,(h-n)\Gamma(h-n+\frac12+i\mu)\Gamma(h-n+\frac12-i\mu)}.
\end{align*}

\subsubsection{pseudo virtual states}

The energy spectrum is depicted in the first graphic
($\mathcal{E}_n(\bm{\lambda})<0\ (n\geq 0)\Leftrightarrow n>2h$) of Fig. 1.
The overshoot eigenfunctions  in region (b) provide the new type of
pseudo virtual state wavefunctions for $\text{v}>2h$:
\begin{equation}
  \tilde{\phi}^{\text{os}}_{\text{v}}(x;\bm{\lambda})
  =\phi_{\text{v}}(x;\bm{\lambda}),\quad
  \tilde{\mathcal{E}}^{\text{os}}_{\text{v}}(\bm{\lambda})
  =\text{v}(2h-\text{v})\quad(\text{v}>2h).
\end{equation}
In the absence of virtual state wavefunctions, shape-invariant rational
extensions of the hyperbolic symmetric top $\II$ potential are not possible.

\subsection{Kepler problem in hyperbolic space (Kh)}
\label{sec:Kh}

This potential is also called Eckart potential.
It has finitely many discrete eigenstates
$0\le n\le n_\text{max}(\bm{\lambda})=[\sqrt{\mu}-g]'$ in the specified
parameter range:
\begin{align*}
  &\bm{\lambda}=(g,\mu),\quad\bm{\delta}=(1,0),\quad
  0<x<\infty,\quad \sqrt{\mu}>g>\frac12,\\
  &w(x;\bm{\lambda})=g\log\sinh x-\frac{\mu}{g}x,\quad
  U(x;\bm{\lambda})=\frac{g(g-1)}{\sinh^2x}
  -2\mu\coth x+g^2+\frac{\mu^2}{g^2},\\
  &\mathcal{E}_n(\bm{\lambda})=g^2-(g+n)^2
  +\frac{\mu^2}{g^2}-\frac{\mu^2}{(g+n)^2},\quad\eta(x)=\coth x,\n
  &f_n(\bm{\lambda})=\frac{\mu^2-g^2(g+n)^2}{g(g+n)^2},\quad
  b_{n-1}(\bm{\lambda})=\frac{n(2g+n)}{g},\\
  &\phi_n(x;\bm{\lambda})
  =e^{-\frac{\mu}{g+n}x}(\sinh x)^{g+n}P_n\bigl(\eta(x);\bm{\lambda}\bigr),
  \quad
  \phi_0(x;\bm{\lambda})
  =e^{-\frac{\mu}{g}x}(\sinh x)^{g},\n
  &P_n(\eta;\bm{\lambda})=P_n^{(\alpha_n,\beta_n)}(\eta),\quad
  \alpha_n=-g-n+\frac{\mu}{g+n},\ \ \beta_n=-g-n-\frac{\mu}{g+n},\\
  &h_n(\bm{\lambda})=\frac{(g+n)\Gamma(1-g+\frac{\mu}{g+n})\Gamma(2g+n)}
  {2^{2g+2n}n!\,\bigl(\frac{\mu^2}{(g+n)^2}-(g+n)^2\bigr)
  \Gamma(g+\frac{\mu}{g+n})}.
\end{align*}
The discrete symmetry and the pseudo virtual state wavefunctions are:
\begin{align}
  &\mathcal{H}(\bm{\lambda})
  =\mathcal{H}\bigl(\mathfrak{t}(\bm{\lambda})\bigr)
  +\mathcal{E}_{-1}(\bm{\lambda}),\quad
  \mathfrak{t}(\bm{\lambda})=(1-g,\mu),\n
  &\tilde{\phi}_{\text{v}}(x;\bm{\lambda})
  =\phi_{\text{v}}\bigl(x;\mathfrak{t}(\bm{\lambda})\bigr)
  =e^{\frac{\mu}{g-\text{v}-1}x}(\sinh x)^{-g+\text{v}+1}
  P_{\text{v}}\bigl(\eta(x);\mathfrak{t}(\bm{\lambda})\bigr),\n
  &\tilde{\mathcal{E}}_{\text{v}}(\bm{\lambda})
  =\mathcal{E}_{\text{v}}\bigl(\mathfrak{t}(\bm{\lambda})\bigr)
  +\mathcal{E}_{-1}(\bm{\lambda})
  =\mathcal{E}_{-\text{v}-1}(\bm{\lambda})
  \quad\bigl(0\le\text{v}<g-1,\ \text{v}>\frac{\mu}{g}+g-1\bigr).
  \label{psenergy}
\end{align}
In the following we restrict to $g>\frac32$.

\subsubsection{virtual states}

The energy spectrum is depicted in the third graphic of Fig. 1,
($\mathcal{E}_n(\bm{\lambda})<0\ (n\geq 0)\Leftrightarrow n>\frac{\mu}{g}-g$,
$\tilde{\mathcal{E}}_{\text{v}}(\bm{\lambda})<0\ (\text{v}\geq 0)
\Leftrightarrow 0\leq\text{v}<g-1$ or $g-1<\text{v}<2g-1$ or
$\text{v}>\frac{\mu}{g}+g-1$).
In this case the discrete symmetry generates type $\II$ virtual state
wavefunctions for $g-1<\text{v}<2g-1$ \cite{quesne5,os29}.
This is the only example, except for (hDPT), that the discrete symmetry
provides the virtual state wavefunctions.
Since this type $\II$ virtual state wavefunction satisfies the boundary
conditions
\begin{equation}
  \partial_x^s\tilde{\phi}_{\text{v}}(x;\bm{\lambda})
  \bigm|_{x=\infty}=0\quad(s=0,1,\ldots),
\end{equation}
a multiple virtual state $\tilde{\phi}_{\text{v}}$ deletion gives
a non-singular Hamiltonian $\mathcal{H}^{[M]}$.
The `shape-invariance' relation is
\begin{align}
  &w_{\mathcal{D}}(x;\bm{\lambda})
  \eqdef\log\biggl|
  \frac{\text{W}[\tilde{\phi}_{d_1},\ldots,\tilde{\phi}_{d_M},\phi_0]
  (x;\bm{\lambda})}
  {\text{W}[\tilde{\phi}_{d_1},\ldots,\tilde{\phi}_{d_M}](x;\bm{\lambda})}
  \biggr|,\n
  &\bigl(\partial_xw_{\mathcal{D}}(x;\bm{\lambda})\bigr)^2
  -\partial_x^2w_{\mathcal{D}}(x;\bm{\lambda})
  =\bigl(\partial_xw_{\mathcal{D}_+}(x;\bm{\lambda}+\bm{\delta})\bigr)^2
  +\partial_x^2w_{\mathcal{D}_+}(x;\bm{\lambda}+\bm{\delta})
  +\mathcal{E}_1(\bm{\lambda}),
  \label{wDsi+}
\end{align}
where $\mathcal{D}_+\eqdef\{d_1+1,d_2+1,\ldots,d_M+1\}$ \cite{os29}.

The overshoot eigenfunctions in region (b) provide type $\I$ virtual
state wavefunctions for $\text{v}>\frac{\mu}{g}-g$ \cite{quesne5}:
\begin{align}
  &\tilde{\phi}^{\text{os}}_{\text{v}}(x;\bm{\lambda})
  =\phi_{\text{v}}(x;\bm{\lambda}),\n
  &\tilde{\mathcal{E}}^{\text{os}}_{\text{v}}(\bm{\lambda})
  =-\text{v}(2g+\text{v})\frac{(g+\text{v}+\frac{\mu}{g})
  (g+\text{v}-\frac{\mu}{g})}{(g+\text{v})^2}
  \quad\bigl(\text{v}>\frac{\mu}{g}-g\bigr).
\end{align}
Since this type $\I$ virtual state wavefunction satisfies the boundary
conditions
\begin{equation}
  \partial_x^s\tilde{\phi}^{\text{os}}_{\text{v}}(x;\bm{\lambda})
  \bigm|_{x=0}=0\quad(s=0,1,\ldots,M-1)\ \ \text{for}\ \ g>M-1,
\end{equation}
a multiple virtual state $\tilde{\phi}^{\text{os}}_{\text{v}}$ deletion gives
a non-singular Hamiltonian $\mathcal{H}^{[M]}$ for $g>M-1$.
It satisfies the `shape-invariance' relation \eqref{wDsi-}.

\subsection{Hyperbolic Darboux-P\"{o}schl-Teller potential (hDPT)}
\label{sec:hDPT}

It has finitely many discrete eigenstates
$0\le n\le n_\text{max}(\bm{\lambda})=[\frac{h-g}{2}]'$ in the specified
parameter range:
\begin{align*}
  &\bm{\lambda}=(g,h),\quad\bm{\delta}=(1,-1),\quad
  0<x<\infty,\quad h>g>\frac12,\\
  &w(x;\bm{\lambda})=g\log\sinh x-h\log\cosh x,\quad
  U(x;\bm{\lambda})=\frac{g(g-1)}{\sinh^2x}
  -\frac{h(h+1)}{\cosh^2 x}+(h-g)^2,\\
  &\mathcal{E}_n(\bm{\lambda})=4n(h-g-n),\quad\eta(x)=\cosh 2x,\quad
  f_n(\bm{\lambda})=2(n+g-h),\quad b_{n-1}(\bm{\lambda})=-2n,\\
  &\phi_n(x;\bm{\lambda})
  =\phi_0(x;\bm{\lambda})P_n\bigl(\eta(x);\bm{\lambda}\bigr),\quad
  \phi_0(x;\bm{\lambda})=(\sinh x)^g(\cosh x)^{-h},\\
  &P_n(\eta;\bm{\lambda})=P_n^{(g-\frac12,-h-\frac12)}(\eta),\quad
  h_n(\bm{\lambda})=\frac{\Gamma(n+g+\frac12)\Gamma(h-g-n+1)}
  {2\,n!\,(h-g-2n)\Gamma(h-n+\frac12)}.
\end{align*}
The eigenvalues can be also expressed as
$\mathcal{E}_n(\bm{\lambda})=4\bigl(\bigl(\tfrac{h-g}{2}\bigr)^2
-\bigl(\tfrac{h-g}{2}-n\bigr)^2\bigr)$.\\
Three types of discrete symmetries are:
\begin{alignat*}{2}
  &\text{type $\I$}:&\quad&\mathcal{H}(\bm{\lambda})
  =\mathcal{H}\bigl(\mathfrak{t}^{\I}(\bm{\lambda})\bigr)
  -(1+2g)(1+2h),\quad
  \mathfrak{t}^{\I}(\bm{\lambda})=(g,-1-h),\\
  &\text{type $\II$}:&\quad&\mathcal{H}(\bm{\lambda})
  =\mathcal{H}\bigl(\mathfrak{t}^{\II}(\bm{\lambda})\bigr)
  -(1-2g)(1-2h),\quad
  \mathfrak{t}^{\II}(\bm{\lambda})=(1-g,h),\\
  &\text{type $\III$}:&\quad&\mathcal{H}(\bm{\lambda})
  =\mathcal{H}\bigl(\mathfrak{t}(\bm{\lambda})\bigr)
  +\mathcal{E}_{-1}(\bm{\lambda}),\quad
  \mathfrak{t}=\mathfrak{t}^{\II}\circ\mathfrak{t}^{\I},
  \ \ \mathfrak{t}(\bm{\lambda})=(1-g,-1-h).
\end{alignat*}
The pseudo virtual state wavefunctions are generated by type $\III$ symmetry
\cite{os29}.\\
In the following we restrict to $g>\frac32$.

\subsubsection{virtual states}

The energy spectrum is depicted in the first graphic
($\mathcal{E}_n(\bm{\lambda})<0\ (n\geq 0)\Leftrightarrow n>h-g$) of Fig. 1.
The overshoot eigenfunctions in region (b) provide type $\I$ virtual state
wavefunctions for $\text{v}>h-g$:
\begin{equation}
  \tilde{\phi}^{\text{os}}_{\text{v}}(x;\bm{\lambda})
  =\phi_{\text{v}}(x;\bm{\lambda}),\quad
  \tilde{\mathcal{E}}^{\text{os}}_{\text{v}}(\bm{\lambda})
  =-4\text{v}(g-h+\text{v})\quad(\text{v}>h-g).
\end{equation}
Since this type $\I$ virtual state wavefunction satisfies the boundary
conditions
\begin{equation}
  \partial_x^s\tilde{\phi}^{\text{os}}_{\text{v}}(x;\bm{\lambda})
  \bigm|_{x=0}=0\quad(s=0,1,\ldots,M-1)\ \ \text{for}\ \ g>M-1,
\end{equation}
a multiple virtual state $\tilde{\phi}^{\text{os}}_{\text{v}}$ deletion gives
a non-singular Hamiltonian $\mathcal{H}^{[M]}$ for $g>M-1$.
It satisfies the `shape-invariance' relation \eqref{wDsi-}.

In the rest of this subsection we present various formulas for the type $\I$
and type $\II$ virtual states which are generated by the discrete symmetries
of the Hamiltonian \cite{os16,hos,os29}:
\begin{align}
  \tilde{\phi}^{\I}_{\text{v}}(x;\bm{\lambda})
  &=\phi_{\text{v}}\bigl(x;\mathfrak{t}^{\I}(\bm{\lambda})\bigr)
  =(\sinh x)^g(\cosh x)^{h+1}
  P_{\text{v}}\bigl(\eta(x);\mathfrak{t}^{\I}(\bm{\lambda})\bigr)
  \quad(\text{v}\geq 0),\n
  \tilde{\mathcal{E}}^{\I}_{\text{v}}(\bm{\lambda})
  &=\mathcal{E}_{\text{v}}\bigl(\mathfrak{t}^{\I}(\bm{\lambda})\bigr)
  -(1+2g)(1+2h)
  =-(2\text{v}+1+2g)(2\text{v}+1+2h),\\
  \tilde{\phi}^{\II}_{\text{v}}(x;\bm{\lambda})
  &=\phi_{\text{v}}\bigl(x;\mathfrak{t}^{\II}(\bm{\lambda})\bigr)
  =(\sinh x)^{1-g}(\cosh x)^{-h}
  P_{\text{v}}\bigl(\eta(x);\mathfrak{t}^{\II}(\bm{\lambda})\bigr)
  \quad(0\leq\text{v}<g-\frac12),\n
  \tilde{\mathcal{E}}^{\II}_{\text{v}}(\bm{\lambda})
  &=\mathcal{E}_{\text{v}}\bigl(\mathfrak{t}^{\II}(\bm{\lambda})\bigr)
  -(1-2g)(1-2h)
  =-(2\text{v}+1-2g)(2\text{v}+1-2h).
\end{align}
Since these virtual state wavefunctions satisfy the boundary conditions
\begin{align}
  &\partial_x^s\tilde{\phi}^{\I}_{\text{v}}(x;\bm{\lambda})
  \bigm|_{x=0}=0\quad(s=0,1,\ldots,M-1)\ \ \text{for}\ \ g>M-1,\\
  &\partial_x^s\tilde{\phi}^{\II}_{\text{v}}(x;\bm{\lambda})
  \bigm|_{x=\infty}=0\quad(s=0,1,\ldots,),
\end{align}
a multiple virtual state $\tilde{\phi}^{\I}_{\text{v}}$
(or $\tilde{\phi}^{\II}_{\text{v}}$) deletion gives a non-singular Hamiltonian
$\mathcal{H}^{[M]}$ ($g>M-1$ for type $\I$).
For a mixed deletion of type $\I$ and $\II$,
\begin{equation*}
  \mathcal{D}=\{d_1,\ldots,d_M\}=\{d^{\I}_1,\ldots,d^{\I}_{M_{\I}},
  d^{\II}_1,\ldots,d^{\II}_{M_{\II}}\}\ \ (M=M_{\I}+M_{\II}),
\end{equation*}
the Wronskian formulas are
\begin{align}
  &\text{W}[\tilde{\phi}_{d_1},\tilde{\phi}_{d_2},\ldots,\tilde{\phi}_{d_M}]
  (x;\bm{\lambda})
  =A_{\mathcal{D}}(x;\bm{\lambda})
  \Xi_{\mathcal{D}}\bigl(\eta(x);\bm{\lambda}\bigr),
  \label{WphitD12}\\
  &\text{W}[\tilde{\phi}_{d_1},\tilde{\phi}_{d_2},\ldots,
  \tilde{\phi}_{d_M},\phi_n](x;\bm{\lambda})
  =A_{\mathcal{D},n}(x;\bm{\lambda})
  P_{\mathcal{D},n}\bigl(\eta(x);\bm{\lambda}\bigr),
  \label{WphitD12n}\\
  &A_{\mathcal{D}}(x;\bm{\lambda})=
  \bigl(\sinh x\bigr)^{g(M_{\I}-M_{\II})+\frac12M_{\I}(M_{\I}-1)
  +\frac12M_{\II}(M_{\II}+1)-M_{\I}M_{\II}}\n
  &\phantom{A_{\mathcal{D}}(x;\bm{\lambda})=}\times
  \bigl(\cosh x\bigr)^{-h(M_{\II}-M_{\I})+\frac12M_{\I}(M_{\I}+1)
  +\frac12M_{\II}(M_{\II}-1)-M_{\I}M_{\II}}
  \times\cF^{-\frac12M(M-1)},\\
  &A_{\mathcal{D},n}(x;\bm{\lambda})=
  \bigl(\sinh x\bigr)^{g(M_{\I}-M_{\II}+1)+\frac12M_{\I}(M_{\I}+1)
  +\frac12M_{\II}(M_{\II}-1)-M_{\I}M_{\II}}\n
  &\phantom{A_{\mathcal{D},n}(x;\bm{\lambda})=}\times
  \bigl(\cosh x\bigr)^{-h(M_{\II}-M_{\I}+1)+\frac12M_{\I}(M_{\I}-1)
  +\frac12M_{\II}(M_{\II}+1)-M_{\I}M_{\II}}
  \times\cF^{-\frac12M(M+1)}.
\end{align}
Here $\Xi_{\mathcal{D}}(\eta;\bm{\lambda})$ and
$P_{\mathcal{D},n}(\eta;\bm{\lambda})$ are polynomials in $\eta$ and
their degrees are generically $\ell'_{\mathcal{D}}$ and
$\ell'_{\mathcal{D}}+n$ respectively, where $\ell'_{\mathcal{D}}$ is
\begin{equation}
  \ell'_{\mathcal{D}}=\sum_{j=1}^{M_{\I}}d^{\I}_j
  +\sum_{j=1}^{M_{\II}}d^{\II}_j
  -\frac12M_{\I}(M_{\I}-1)-\frac12M_{\II}(M_{\II}-1)+M_{\I}M_{\II}.
  \label{ellD'}
\end{equation}
The ratio of the Wronskians becomes
\begin{equation}
  \frac{\text{W}[\tilde{\phi}_{d_1},\ldots,\tilde{\phi}_{d_M},\phi_n]
  (x;\bm{\lambda})}
  {\text{W}[\tilde{\phi}_{d_1},\ldots,\tilde{\phi}_{d_M}](x;\bm{\lambda})}
  =\frac{P_{\mathcal{D},n}\bigl(\eta(x);\bm{\lambda}\bigr)}
  {\Xi_{\mathcal{D}}\bigl(\eta(x);\bm{\lambda}\bigr)}
  \times
  \phi_0(x;\bm{\lambda}+M_{\I}\tilde{\bm{\delta}}^{\I}
  +M_{\II}\tilde{\bm{\delta}}^{\II}),
  \label{WphitD12n/WphitD12}
\end{equation}
where $\tilde{\bm{\delta}}^{\I}=(1,1)$ and $\tilde{\bm{\delta}}^{\II}=(-1,-1)$.
The lowest degree polynomial $P_{\mathcal{D},0}(\eta;\bm{\lambda})$ is
related to the denominator polynomial $\Xi_{\mathcal{D}}$ with shifted
parameters
\begin{equation}
  P_{\mathcal{D},0}(\eta;\bm{\lambda})
  \propto\Xi_{\mathcal{D}}(\eta;\bm{\lambda}+\bm{\delta}).
\end{equation}
For a proper choice of the parameter range,
the Hamiltonian $\mathcal{H}^{[M]}$ is non-singular as in the trigonometric
DPT case \cite{os25}.
The ordinary shape-invariance relation holds:
\begin{align}
  &w_{\mathcal{D}}(x;\bm{\lambda})
  =\log\biggl|
  \frac{\text{W}[\tilde{\phi}_{d_1},\ldots,\tilde{\phi}_{d_M},\phi_0]
  (x;\bm{\lambda})}
  {\text{W}[\tilde{\phi}_{d_1},\ldots,\tilde{\phi}_{d_M}](x;\bm{\lambda})}
  \biggr|,\n
  &\bigl(\partial_xw_{\mathcal{D}}(x;\bm{\lambda})\bigr)^2
  -\partial_x^2w_{\mathcal{D}}(x;\bm{\lambda})
  =\bigl(\partial_xw_{\mathcal{D}}(x;\bm{\lambda}+\bm{\delta})\bigr)^2
  +\partial_x^2w_{\mathcal{D}}(x;\bm{\lambda}+\bm{\delta})
  +\mathcal{E}_1(\bm{\lambda}).
  \label{wDsi}
\end{align}

\subsection{New pseudo virtual states with half integer coupling constant}
\label{sec:halfint}

A Darboux transformation in terms of a pseudo virtual state
$(\tilde{\phi}_{\text v}(x),\tilde{\mathcal E}_{\text v})$ creates
a new eigenstate with the energy $\tilde{\mathcal E}_{\text v}$.
For the ordinary pseudo virtual states created by the twisting of parameters,
the energy is always equal to that of the original `eigenstate' with a
negative degree
$\tilde{\mathcal E}_{\text v}=\mathcal{E}_{-\text{v}-1}$ \eqref{psenergy},
as emphasised in \cite{os29}.
For the new type of pseudo virtual states found in (s) \S\ref{sec:s},
(RM) \S\ref{sec:RM} and (hst) \S\ref{sec:hst}, their energies do not
coincide with the negative energy levels of the original system in general.
However it happens for half integer coupling constant cases. 
In these cases the possibility arises that the Darboux-Crum transformations
in terms of multiple pseudo virtual states are equivalent to the Krein-Adler
transformations in terms of multiple eigenstates with shifted parameters,
like those explained in \cite{os29}.
In fact, this happens for (s) with a half odd integer $h$ and for (RM) with
a half integer $h$.

In (RM), the energy function $\mathcal{E}_n(\bm{\lambda})$ ($n\in\mathbb{R}$)
satisfies $\mathcal{E}_n(\bm{\lambda})=\mathcal{E}_{2h-n}(\bm{\lambda})$.
In the following we assume that the coupling constant $h$ is a half integer,
$2h\in\mathbb{Z}_{>0}$.
Then, for integers $\text{v}$ ($\text{v}>2h$) and
$\text{v}'=\text{v}-2h-1\geq 0$, we have $\mathcal{E}_{\text{v}}(\bm{\lambda})
=\mathcal{E}_{-\text{v}'-1}(\bm{\lambda})$, namely
$\tilde{\mathcal{E}}^{\text{os}}_{\text{v}}(\bm{\lambda})
=\mathcal{E}_{-\text{v}'-1}(\bm{\lambda})$, which resembles
$\tilde{\mathcal{E}}_{\text{v}}(\bm{\lambda})
=\mathcal{E}_{-\text{v}-1}(\bm{\lambda})$ in \cite{os29}.
Moreover, it is easy to see that the degree of the polynomial is reduced
accordingly,
\begin{equation}
  \text{deg}\bigl(P_\text{v}(\eta;\bm{\lambda})\bigr)
  =\text{v}-2h-1=\text{v}'\quad(\text{v}>2h).
  \label{degred}
\end{equation}
Let $\mathcal{D}\eqdef\{d_1,d_2,\ldots,d_M\}$ ($d_j>2h$)
be a set of distinct non-negative integers and fix an integer
$N\geq\text{max}(\mathcal{D})-1$.
Let us define a set of distinct non-negative integers
$\bar{\mathcal{D}}=\{0,1,\ldots,N\}\backslash
\{\bar{d}'_1,\bar{d}'_2,\ldots,\bar{d}'_M\}$
together with the shifted parameters $\bar{\bm{\lambda}}$:
\begin{align}
  &\bar{\mathcal{D}}\eqdef\{0,1,\ldots,\breve{\bar{d}}'_1,\ldots,
  \breve{\bar{d}}'_2,\ldots,\breve{\bar{d}}'_M,\ldots,N\}
  =\{e_1,e_2,\ldots,e_{N+1-M}\},\n
  &d'_j\eqdef d_j-2h-1\geq 0,\quad\bar{d}'_j\eqdef N-d'_j\geq 0,\quad
  \bar{\bm{\lambda}}\eqdef \bm{\lambda}-(N+1)\bm{\delta}.
  \label{barD}
\end{align}
Then we have
\begin{equation}
  \bar{\Xi}_{\mathcal{D}}(\eta;\bm{\lambda})\propto
  \bar{\Xi}_{\bar{\mathcal{D}}}(\eta;\bar{\bm{\lambda}}),
  \label{detiden}
\end{equation}
whose degree is $\ell_{\bar{\mathcal{D}}}=\ell_{\mathcal{D}}-(2h+1)M
=\sum_{j=1}^Md'_j-\frac12M(M-1)<\ell_{\mathcal{D}}$.
The relation \eqref{detiden} implies the equivalence of the two deformed
systems;
the Darboux-Crum transformations in terms of $\mathcal{D}$ pseudo virtual
states $\tilde{\phi}^{\text{os}}_{d_j}$ and
the Krein-Adler transformations in terms of $\bar{\mathcal{D}}$ eigenstates
$\phi_{e_j}$ with shifted parameters $\bar{\bm{\lambda}}$,
as explained in detail in \cite{os29}.
The situation is similar for (s) with a half odd integer $h$. 
The degree reduction of the polynomial is realised by factorisation:
\begin{equation*}
  P_{\text{v}}(\eta;\bm{\lambda})=(1+\eta^2)^{h+\frac12}
  \times\bigl(\text{a polynomial of degree $\text{v}-2h-1$ in $\eta$}\bigr)
  \quad(\text{v}>2h), 
\end{equation*}
and the above equivalence holds with a slight modification on the left hand
side:
\begin{equation}
  (1+\eta^2)^{-(h+\frac12)M}\bar{\Xi}_{\mathcal{D}}(\eta;\bm{\lambda})
  \propto\bar{\Xi}_{\bar{\mathcal{D}}}(\eta;\bar{\bm{\lambda}}).
\end{equation}
For an integer $h$ ($h=m$), the polynomial $P_{\text{v}}(\eta;\bm{\lambda})$
($\text{v}>2h$) vanishes but the limit
$\displaystyle\lim_{h\to m}(h-m)^{-1}P_{\text{v}}(\eta;\bm{\lambda})$ exists
and its degree is $\text{v}$.
Thus it is not divisible by a non-polynomial factor $(1+\eta^2)^{m+\frac12}$.
The degree reduction \eqref{degred} does not occur for (hst), too.
Thus the above equivalence does not hold for (s) with an integer $h$ and (hst).

\section{Summary and Comments}
\label{summary}

For iso-spectral rational extensions of shape-invariant potentials in
one dimensional quantum mechanics, virtual state wavefunctions are essential.
For the radial oscillator potentials and the Darboux-P\"oschl-Teller
potentials and various examples from the discrete quantum mechanics,
the virtual state wavefunctions are obtained by twisting parameters
based on discrete symmetries of the Hamiltonian
\cite{os25,os26,os27,os16,os17,os19,os23,os29}.
For six shape-invariant potentials having finitely many discrete
eigenstates, the concept of overshoot eigenfunctions is introduced for
searching new types of virtual and pseudo virtual wavefunctions.
The new type of virtual state wavefunctions exist for (M), (RM), (Kh)
\cite{quesne5} and (hDPT).
As a byproduct, a new type of pseudo virtual state wavefunctions are
discovered for (s), (RM) and (hst).

With the present paper, our research project on the rational extensions of
various exactly solvable potentials in one dimension is almost complete
\cite{os25,os16,os18,os19,hos,stz,os21,gos,os29}.
Establishing parallel results in one dimensional discrete quantum mechanics
(those having the Wilson, Askey-Wilson, ($q$-)Racah polynomials, etc as the
main part of the eigenfunctions) is an immediate task.
An interesting challenge is  rational extensions of exactly solvable
multi-dimensional quantum mechanical models, {\em e.g.} the Calogero-Moser,
the Ruijsenaars-Schneider-van Diejen models, etc.

\section*{Acknowledgements}
We thank A.\,Khare for useful discussion at the early stage of investigation.
R.\,S. is supported in part by Grant-in-Aid for Scientific Research
from the Ministry of Education, Culture, Sports, Science and Technology
(MEXT), No.22540186.


\end{document}